# Underwater Acoustic Detection of Ultra High Energy Neutrinos

V. Niess and V. Bertin

*Centre de Physique des Particules de Marseille, CNRS/IN2P3*
*Université de la Méditerranée Aix-Marseille II, Marseille, France*

## Abstract

We investigate the acoustic detection method of $10^{18\text{-}20}$ eV neutrinos in a Mediterranean Sea environment. The acoustic signal is re-evaluated according to dedicated cascade simulations and a complex phase dependant absorption model, and compared to previous studies. We detail the evolution of the acoustic signal as function of the primary shower characteristics and of the acoustic propagation range. The effective volume of detection for a single hydrophone is given taking into account the limitations due to sea bed and surface boundaries as well as refraction effects. For this 'benchmark detector' we present sensitivity limits to astrophysical neutrino fluxes, from which sensitivity bounds for a larger acoustic detector can be derived. Results suggest that with a limited instrumentation the acoustic method would be more efficient at extreme energies, above $10^{20}$ eV.

**Keywords**
cosmic neutrinos; underwater particle showers; underwater acoustics; neutrino detectors.

## 1) INTRODUCTION

The search for ultra high energy (UHE, $10^{18\text{-}20}$ eV) neutrinos is strongly motivated by the observation of cosmic rays. Since the first comprehensive observations by Auger in 1938, it has been observed that by extending the coverage of detectors, increasingly higher energy events have been detected. Presently, the spectrum of these outer-space particles is known to extend up to $10^{20}$ eV, and maybe even higher. In particular, there is an on-going discussion on the existence, or not, of the GZK [1] cut-off on the cosmic rays spectrum. Because neutrinos interact so weakly, their observation at UHE would bring insight on the origin and nature of these UHE cosmic rays. If the GZK cut-off is relevant, it could be studied from the observation of neutrinos which will be produced as secondaries from pion disintegrations. This would signal the existence of extremely high energies cosmic particles (EHE, $10^{20+}$ eV). Furthermore, neutrinos are the only known candidates that would allow astrophysical observation at these extreme energies, above the GZK cut-off.

Because expected fluxes are low, and because neutrinos interact so weakly, huge target volumes are required to efficiently detect UHE cosmic neutrinos. Hence, it is interesting to exploit natural dense media such as the sea or polar ice. In this work we investigate a particular acoustic detection method and its application to the observation of UHE neutrinos. An acoustic emission develops from the local heating resulting from the energy deposition of underwater cascades induced by neutrino interactions in sea water. In the framework of the Standard Model, UHE neutrinos are expected to interact by deep inelastic scattering with the nuclei of water molecules, resulting in hadronic fragmentation and, in the case of a charged current interactions (CC), in a charged lepton. Following Gandhi *et al.* [2], the interaction is almost elastic, with a 70% probability of producing a charged lepton, taking away 80% of the energy of the primary neutrino. Nevertheless, when a $\mu$ or $\tau$ lepton is produced, the total extent of the lepton travel path is expected to be large, a few tens to a hundred of kilometres at UHE [3], as compared to the available water depth of a few kilometres. The 'golden' event for acoustic detection is a $\nu_e$ charged current interaction, where all the energy of the primary neutrino is guaranteed to be dumped into cascades with a relatively short extent ($\approx 10 - 100$ m). In the case of a neutral current interaction (NC), on average, only 20% of the initial energy of the neutrino is available for acoustic detection, via a hadronic shower. In this work we focus on these two latter specific cases.

This paper follows a work [4] conducted at Centre de Physique des Particules de Marseille (CPPM) in association with the ANTARES neutrino telescope [5], which is based on a detection technique with Cerenkov light. Consequently, some aspects of this work are strongly related to the Mediterranean Sea and the ANTARES site location. This paper highlights some of the important points for the acoustic technique. The first part of the paper reports on acoustic signal simulations and studies. The second part is related to detection performances estimates.

## 2) SIGNAL STUDIES

### 2-1) The thermo-acoustic emission mechanism

A cascade of relativistic particles develops in the interaction medium almost at the speed of light in vacuum leading to a fast local heating of a macroscopic volume of ionised water. As first predicted by Askariyan [6] and confirmed experimentally with proton beam experiments by Sulak *et al.* [7], the heated water volume, 'instantaneously' deviated from thermodynamic equilibrium, relaxes first by an acoustic coupling mechanism. This results in the emission of an impulse of pressure, *p*, described by a wave equation with a so called 'thermo-acoustic' (TA) source term, given as:

$$\rho_0 \vec{\nabla} \cdot (\frac{1}{\rho_0} \vec{\nabla} p) - \frac{1}{c_s^2} \frac{\partial^2 p}{\partial t^2} = -\frac{\alpha}{C_p} \frac{\partial^2 q}{\partial t^2} \qquad (1)$$

where *q* is the energy deposition density along the cascade. The acoustic emission depends on sea water characteristics: its density $\rho_0$, the adiabatic and non-dispersive speed of sound $c_s$, the thermal expansion coefficient $\alpha$ and the heat capacity $C_p$. These four parameters depend on the water temperature, its hydrostatic pressure and its salinity. Consequently, the efficiency of the thermo-acoustic conversion mechanism varies over the water volume. On kilometric ranges, sea water is mostly a vertically stratified medium with a depth dependent sound speed profile. This leads to refraction effects for the wave-front discussed further in section 3-3.

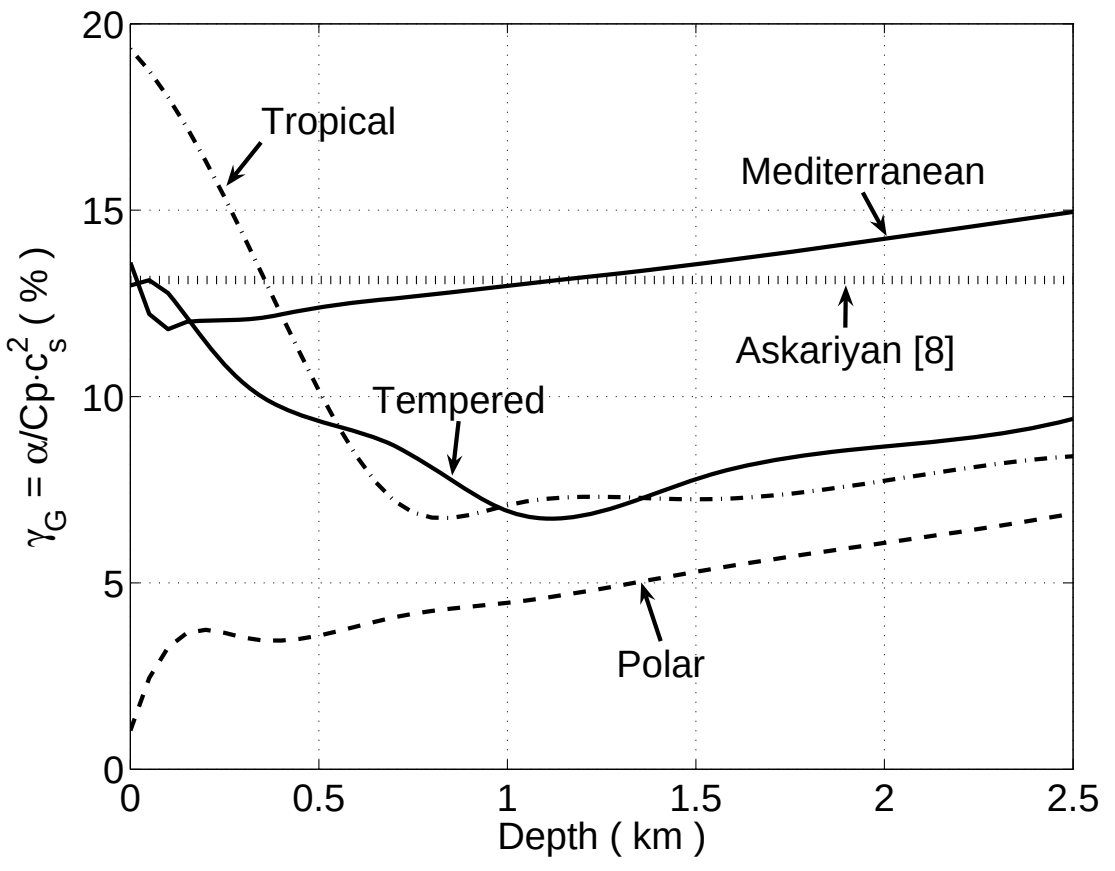

Figure 1 : Relative strength of the acoustic signal for different oceans and seas. The strength of the signal is expressed as Gruneisen parameter, $\gamma_G$, as a function of depth. The various lines are polar areas (dashed), oceans in tropical (dash-dotted) and tempered (solid, bottom line) areas, and the Mediterranean Sea (solid, upper line). The dotted horizontal line indicates the value used by Askariyan *et al.* [8] as well as in this paper.

The temperature and salinity conditions also vary from one sea to another. A *figure of merit* for the acoustic pulse relative intensity is given by the dimensionless Gruneisen parameter $\gamma_G = \alpha c_s^2 / C_p$. The signal strength increases with salinity, temperature and depth. Examples are given on figure 1 for different oceans and seas. The Mediterranean Sea, which is an 'old' and 'closed' sea, has a high salinity with an extraordinary warm temperature at large depth, close to 13°C. This results in an enhancement by a factor of 2 of the strength of the acoustic emission, as compared to the oceans. In particular it is relevant to recall that in fresh water the thermal expansion coefficient crosses zero at about 4°C, which is close to the average temperature of most kilometric deep 'natural' water volumes.

For this study we assumed a constant value for the Gruneisen parameter. For the purpose of comparison with a previous study by Askariyan *et al.* [8] we take $C_p = 3.6 \cdot 10^3\ \mathrm{J \cdot kg^{-1} \cdot K^{-1}}$ and $\alpha = 2.1 \cdot 10^{-4}\ \mathrm{K^{-1}}$. These values are also representative of the Mediterranean Sea at a 1 km depth. For comparison a similar strength is reached only in the deepest part of the oceans at 6 km depths.

### 2-2) Energy deposition in underwater showers

As a first step for the computation of our acoustic signal, it is necessary to describe the energy deposition along the cascades, hence to study the development of UHE underwater showers. Due to the lack of experimental data in the UHE range for dense media, such studies rely mostly on Monte-Carlo simulations and/or extrapolations from lower energy experimental data. Previous studies of this energy deposition in water have been given for water by Askaryian *et al.* [8], Stanev *et al.* [9], more recently by Dedenko *et al.* [10] and, in ice, by Alvarez-Muniz and Zas [11]. Results from [11] have been adapted to water by Lehtinen *et al.* [12]. It was seen from the work of Alvarez-Muniz and Zas that whereas the Landau-Pomeranchuk-Migdal (LPM) effect [13,14] is crucial for UHE EM showers, it is less relevant for hadronic initiated showers.

Assuming axial symmetry by rotation around the primary direction for the energy deposition, it is relevant to split the energy deposition, *q*, over two distribution functions as follow :

$$q(\rho, z, t) = \frac{1}{2\pi} f_z(z) g_z(\rho, z) H(t - \frac{z}{c_0}) \qquad (2)$$

where *z* is the depth along the shower axis, $\rho$ the lateral distance to the z-axis and *t* the time, starting from the primary interaction. The functions, $f_z$, and, $g_z$, characterise the longitudinal and lateral energy deposition density. *H* is the Heaviside distribution, hence we assume here that the shower develops longitudinally at the speed of light in vacuum, $c_0$. The normalisation conditions are taken as :

$$\int f_z(z)dz = E \quad \text{and} \quad \int \rho g_z(\rho, z)d\rho = 1 \qquad (3)$$

with *E* the total energy dumped into the cascade.

### 2-2-1) Longitudinal distribution of energy deposition

The hadronic initiated part of the shower was described by a deterministic longitudinal distribution parameterised according to the Particle Data Group (PDG) [3] as:

$$f_z(z) = \frac{E}{X_0} b \frac{(bu)^{a-1}\exp(-bu)}{\Gamma(a)}, \quad u = z / X_0 \quad (4)$$

where, $X_0$, is the radiation length, close to 36 cm in sea water, and with, $\Gamma$, the gamma function. Parameters, *a,* and, *b,* depend on the primary particle nature and the energy *E* of the cascade. For the purpose of this study these parameters were tuned according to GEANT4 [15] simulations in the energy range from 100 GeV to 1 PeV, and extrapolated to higher energies. Hadronic interactions in GEANT4 were modelled with the QGSP physics list [15], dedicated to high energy calorimetry on collider experiments. For such a model, the depth of the maximum of deposited energy, given as $z_{max}=(a-1)/b$, increases logarithmically with the energy *E* as $z_{max} \propto D\ln(E)$. The parameter, *D,* was found close to 0.65 for hadronic initiated showers, which is consistent with extended air showers observations (EAS) [16]. Hence the extension of the cascade varies only slightly with the energy. In the UHE range, hadronic initiated cascades are about 10 m long in water.

For the electromagnetic initiated part, the LPM suppression effect is of great importance. A dedicated algorithm was developed to simulate EM showers in water up to $10^{20}$ eV. This algorithm which is described in appendix A relies on a two step scheme. For the Bremsstrahlung and pair production cross sections we used computations from Migdal [14]. It was seen that up to $10^{17}$ eV, the longitudinal distribution of showers is still well described by the mean parameterisation given in equation (4). However, starting from $10^{15}$ eV, values of the parameters *a* and *b* deviate from their GEANT4 extrapolation from lower energies. At this $10^{15}$ eV energy threshold, the effective radiation length of the Bremsstrahlung process differs from the value $X_0$. It begins to increase as a square-root power-law due to the LPM suppression effect [4]. As a result the shower length is extended, as can be seen in figure 2.

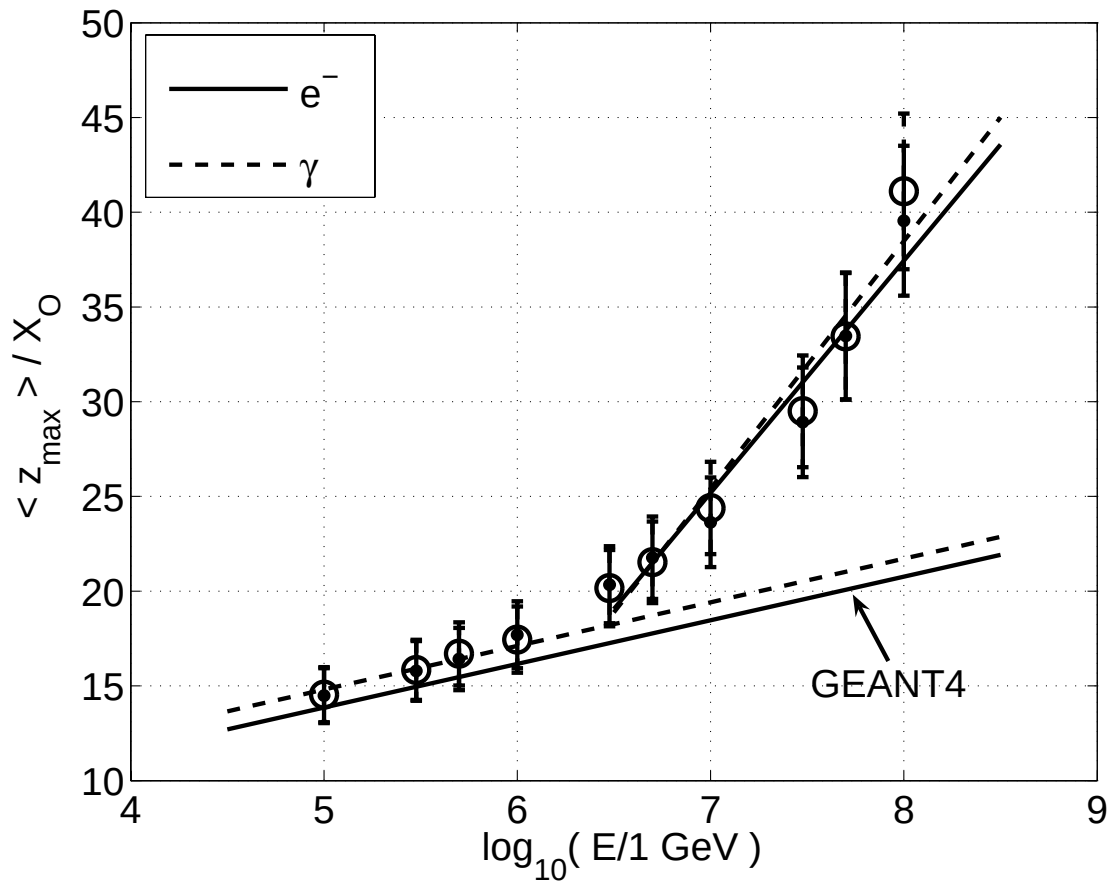

Figure 2: Depth of the maximal of energy deposition as a function of primary energy. Dots ($e^-$) and circles ($\gamma$) corresponds to results of the Monte-Carlo simulation. For both primaries, error bars indicate the standard deviation on a single cascade event. The extrapolation from GEANT4 simulations of EM cascades is indicated.

At UHE the longitudinal energy distribution becomes stochastic. The mean parameterisation given by equation (4) is no longer relevant to describe the shape of the longitudinal distribution. Figure 3 gives some examples to illustrate this effect and so, in the paper, Monte-Carlo simulations are used. The mean depth of the maximum of energy deposition, $<z_{max}>$, increases fast with energy, as a power-law, whereas the normalised mean density falls, also as a power-law; this can be seen on figure 4. Values of exponents are close to 0.5, compatible with the behaviour of the Bremsstrahlung and pair production processes in the LPM deep suppression regime [4].

Though the Midgdal cross sections used for this study have shown agreement with experimental measurements obtained at colliders [17,18] in the sub TeV domain, there are not experimentally constrained in the deep suppression regime reached in UHE EM showers. In particular, precise extrapolation to this regime is subject to discussion, and cross sections might be miss-evaluated by a factor of 2 at UHE [4,19]. Consequently, the extension of LPM cascades could be overestimated by a similar amount. In addition, at extreme energies, Bremsstrahlung and pair production cross sections are suppressed to a few mbarn. Competitive mechanisms may come into play, slowing down the development of the cascade. For example, Fletcher et al. [20] predict cross-sections of a few mbarn at UHE for photo-production of hadrons.

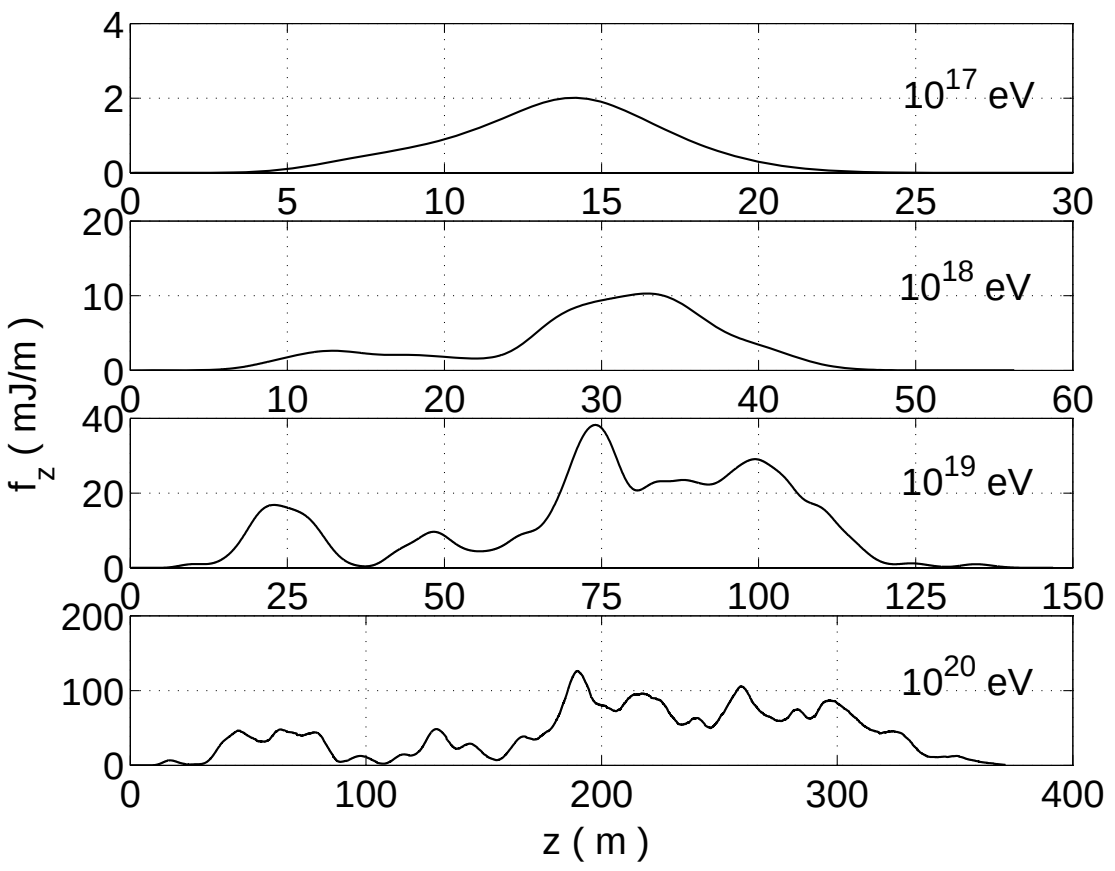

Figure 3: Examples of longitudinal distribution for LPM extended cascades for various energies. The cascade energy is indicated on the upper right part of each plot.

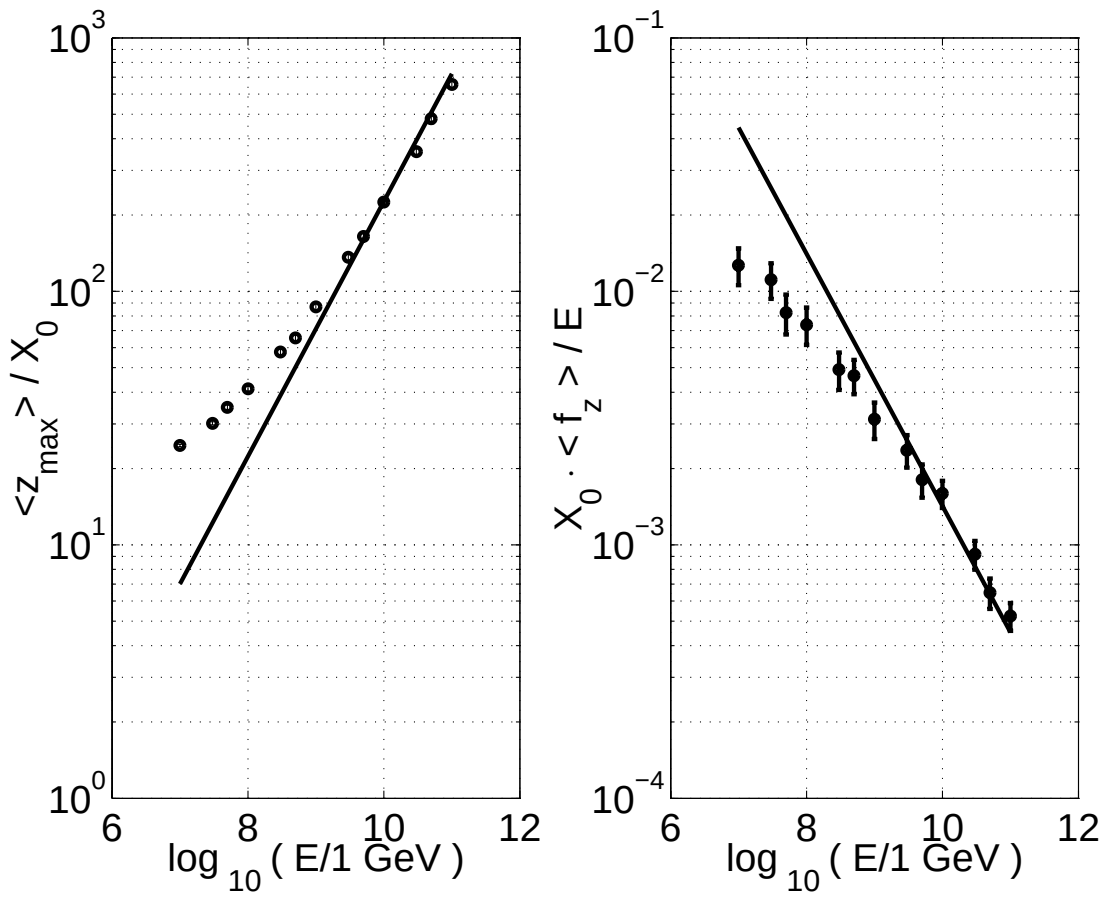

Figure 4: Asymptotic behaviour of LPM cascades. The plot on the left shows the depth of maximal energy deposition. The plot on the right is the mean longitudinal energy deposition density. Density is scaled by the primary energy, *E*, and by the radiation length, $X_0$. Error bars indicate standard deviation on the mean value estimation.

### 2-2-2) Lateral distribution of energy deposition

From the GEANT4 simulations, the lateral distribution of energy deposition $g_z$ was seen to follow different broken power-laws at short and long ranges from the shower axis, with a transition at a radius $\rho_i$ of about 4 cm. The lateral distribution was parameterised as follow :

$$\frac{g_z(\rho,z)}{g_0} = \begin{cases} x^{n1} & \text{if} \quad x \geq 1 \\ x^{n2} & \text{otherwise} \end{cases}, \quad x = \rho_i / \rho \tag{5}$$

and $g_0 = \frac{1}{\rho_i^2} \frac{(2-n_1)(n_2-1)}{n_2 - n_1}$

where the constant $g_0$ is computed from the normalisation condition as given in equation (3). Values of lateral exponents $n_1$, for the core and $n_2$ in the peripheral area, depend on the depth along the shower. They decrease when moving deeper along the cascade, reflecting a broadening of the shower. These results are in qualitative agreement with extended air showers observations and the NKG [21] parameterisation. However, this latter parameterisation failed to reproduce the fast inflexion around radius $\rho_i$ leading to an inconsistent cascade age parameter at the depth of maximum energy deposition, hence we used this broken power-law parameterisation in order to reproduce the core-density behaviour. In addition the inflexion does not occur at the Moliere radius, of 10.8 cm for water according to PDG [3], contrary to what would have been expected from the NKG parameterisation.

From GEANT4 cascades the dependency of the shape of the lateral distribution with the depth along the shower was found to scale with the depth, $z_{max}$, of the maximum density. Once scaled, the core exponent does not depend significantly on the primary energy and on its nature, however, the peripheral exponent does depend on the nature of the primary and on its energy. In the neighbourhood of the maximum density, for EM cascades, the lateral exponents follow a linear law with depth as :

$$\begin{cases} n_1 = 1.66 \pm 0.02 - (0.29 \pm 0.02)(z / z_{max}) \\ n_2 = 4.35 \pm 0.05 - (1.10 \pm 0.04)(z / z_{max}) \end{cases} \tag{6}$$

The value of the core exponent was found to be within 10% agreement with values observed experimentally by ground arrays surveying EAS [16]. However, it should be pointed out that these experimental observations, sample mostly the peripheral area of the lateral distribution of particles. The peripheral exponent is only in 25% agreement with our simulations. The larger disagreement might be explained by the dependency on the primary nature and energy.

For LPM showers it was found that reaching the UHE regime the lateral exponent becomes constant along the shower depth, tending to the value observed at the depth $z_{max}$ of maximum density for lower energy cascades.

### 2-3) Propagation loss

Due to the ‘instantaneous’ energy deposition and the narrow core lateral distribution, the acoustic signal radiated by the shower is dominated by an ultra-sonic frequency content. However, water, and in particular sea water, does not allow the propagation of the highest acoustic frequencies, with an attenuation length decreasing as a frequency squared power-law, hence, especially at large ranges, the shape of the acoustic signal coming from a cascade is strongly modelled by propagation losses.

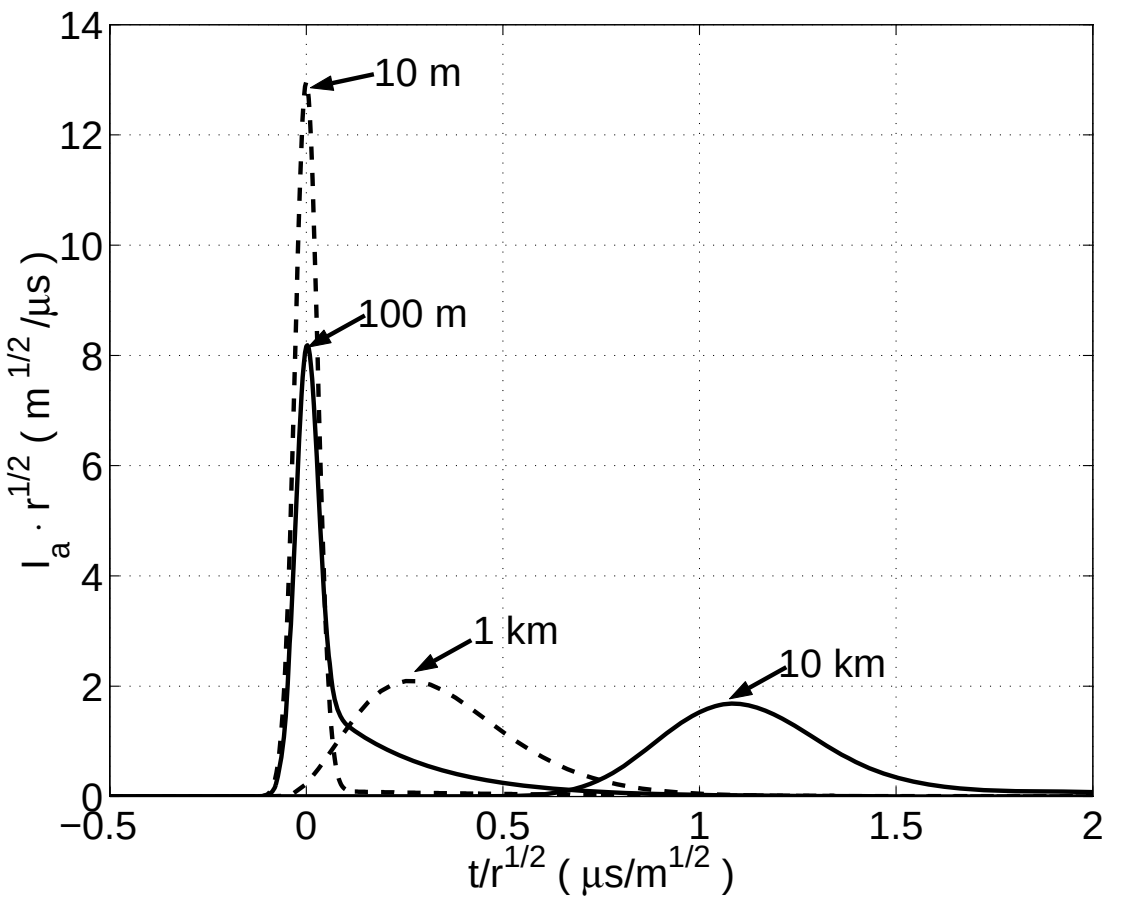

Figure 5: Absorption impulse response at various ranges as indicated in the figure. Impulse response, $I_a$, and time scale are normalized by the square-root of the distance, *r*, for better comparison.

These losses are dominated by viscosity effects at frequencies above 100 kHz and, for sea water, chemical coupling mechanisms at lower frequencies. The dissolution equilibrium of some salts, especially magnesium sulfate ($MgSO_4$), is particularly sensitive to density variations. As a result, a significant part of the acoustic energy of the pressure wave transfers to a displacement of the chemical equilibrium. According to the work of Liebermann [22] the complex absorption coefficient can be approximated as :

$$\tilde{a}(i\omega) = \frac{\omega^2}{\omega_0 c_s} + \left(\frac{1}{r_1}\right)\frac{i\omega}{\omega_1 + i\omega} + \left(\frac{1}{r_2}\right)\frac{i\omega}{\omega_2 + i\omega} \quad (7)$$

where $\omega = 2\pi f$ is the angular frequency. The first term reflects loss from viscosity, whereas the two other terms are the dominant contributions from chemical equilibrium of $MgSO_4$ and boric acid ( $B(OH)_3$ ). Values of these parameters were fitted to Mediterranean Sea conditions, according to results of Francois and Garrison [23]. For this study, the absorption coefficient was approximated as constant over the water volume with parameter values given as:

$$\begin{bmatrix} \omega_0 = 0.79 \cdot 10^{12} \text{ rad/s} & c_s = 1520 \text{ m/s} \\ f_1 = 91.2 \text{ kHz} & r_1 = 157.8 \text{ m} \\ f_2 = 1.31 \text{ kHz} & r_2 = 45.6 \text{ km} \end{bmatrix} \quad (8)$$

The absorption impulse response $I_a(r,t)$ is computed as :

$$I_a(\vec{r},t) = \frac{1}{2\pi}\int_{-\infty}^{+\infty} \exp(-\tilde{a}(i\omega)r)d\omega \quad (9)$$

where *r* is the distance between the observation point and the sonic source. The impulse response is computed numerically, by inverse Fourier transform, from the complex absorption coefficient, $\tilde{a}$, given in equation (7). Results are shown on figure 5. At distances below $r_1 = 157.8$ m propagation loss is dominated by viscosity. The impulse response is characterised by a gaussian diffusion behaviour, with an amplitude falling with the square-root of the distance, and a duration increasing as $\sqrt{r}$. As chemical losses from magnesium sulfate comes into play, the impulse response first exhibits a tail and then begins to decrease in amplitude, with an increasing characteristic duration. At ranges of 1 km and more, the response still exhibits a gaussian diffusion shape, but with an amplitude lower by one order of magnitude, as compared to short ranges, and a consequently longer duration.

## 2-4) Acoustic signal computation

An integral expression of the pressure field is given by the Green's function associated to equation (1). In a general case this is not a simple problem considering that the sound velocity and water density vary with depth. However, one should recall that we are interested in an impulse like pressure signal dominated by a high frequency content. Under these conditions, in the frequency domain, phase corrections are the dominant effects. Hence, neglecting (de-)focusing effects from refraction the pressure field is given by a 3-dimensional integral over the energy deposition as follow :

$$p(\vec{r},t) = \frac{\alpha}{4\pi C_p}\frac{\partial}{\partial t}\int \frac{\delta(t-\tau(\vec{r},\vec{r}'))}{|\vec{r}-\vec{r}'|} q(\vec{r}')d^3\vec{r}' \quad (10)$$

where $\tau(\vec{r},\vec{r}')$ is the propagation time from $\vec{r}$ to $\vec{r}'$ given by a ray tracing model. Furthermore we neglected variations of thermal expansion and specific heat capacity over the volume of the energy deposition. It can be shown [4] that for a linear gradient of sound velocity, as it is the case in the Mediterranean Sea, equation (10) is valid to an accuracy of a few percent.

The integral given by equation (10) is reduced to 1-dimensional integral by exploiting the causality property and the axial symmetry of the cascade energy distribution. Further details are given in appendix B. It is also worth pointing out that due to the impulse-like nature of the signal, propagation losses for extended sources are given over the full range by a simple convolution with the absorption impulse response. Using the cylindrical coordinates introduced in part 2-2 for the energy deposition, it goes as:

$$p_a(\rho,z,t) = p(\rho,z,t) \otimes I_a(\rho,t) \quad (11)$$

where $p_a$ is the 'absorbed' pressure field and $I_a$ the absorption impulse response defined in equation (9). Complications due to variations in the travel path from different points of the extended source to the observer are not relevant, because the acoustic emission has a local origin. Hence, in the absorption impulse response $I_a$, one can approximate the propagation range, $r$, with the lateral distance, $\rho$, to the cascade axis. In addition, due to this local sound emission origin, the delay induced by the development of the shower at the finite speed of light $c_0$ is taken into account by a simple time translation along the cascade depth, as :

$$p_{c_0}(\rho, z, t) = p(\rho, z, t - z/c_0) \qquad (12)$$

**2-5) Acoustic signal characteristics**

The acoustic pressure field has been computed for both LPM extended EM cascades and hadronic compact energy depositions and for observation ranges varying from 1 m to 10 km. For the lateral distribution we used a core power-law, as given by equation (5), with a 'hard' numerical cut at a distance of 200 μm (2% of Moliere radius) from the shower axis. This differs from previous work [8,24] where exponential or Lorentzian laws were used. A core power-law behaviour to such low distances from the cascade axis, is supported by experimental microscopic observations of showers in sandwiches of lead plate absorbers and emulsions [25].

The general characteristics of the acoustic signal are as follows. Due to the strong anisotropy of the source (the cascade extends from tens to hundred metres long, but laterally the energy deposition is confined to a few centimetres diameter) the emission is strongly directive, confined to directions orthogonal to the shower axis. Due to the rapid energy deposition, the signal exhibits a bipolar impulse structure with a short and 'strong' leading compression peak, followed by a longer but weaker rarefaction, or relaxation peak. This is sketched on figure 6. The pulse like time structure results in a broadband frequency spectrum, in creasing as ~*f*, hence dominated by an ultra-sonic frequency content. The spectrum increase with frequency is limited by two main factors : coherency lost due to the lateral spread of the energy deposition and scattering from sound absorption.

Characteristics of the signal vary with the extension of the cascade and the range and therefore it is convenient to characterise the signal by its peak to peak duration, $\Delta t$, its half peak to peak amplitude, $A_p$, and a symmetry factor, *R/C*, defined as the ratio of the amplitudes of the rarefaction peak to the compression one. This is shown on figure 6. In addition we define the arrival time of the signal, $t_a$, as the intermediary time between detection of the compression and rarefaction peaks.

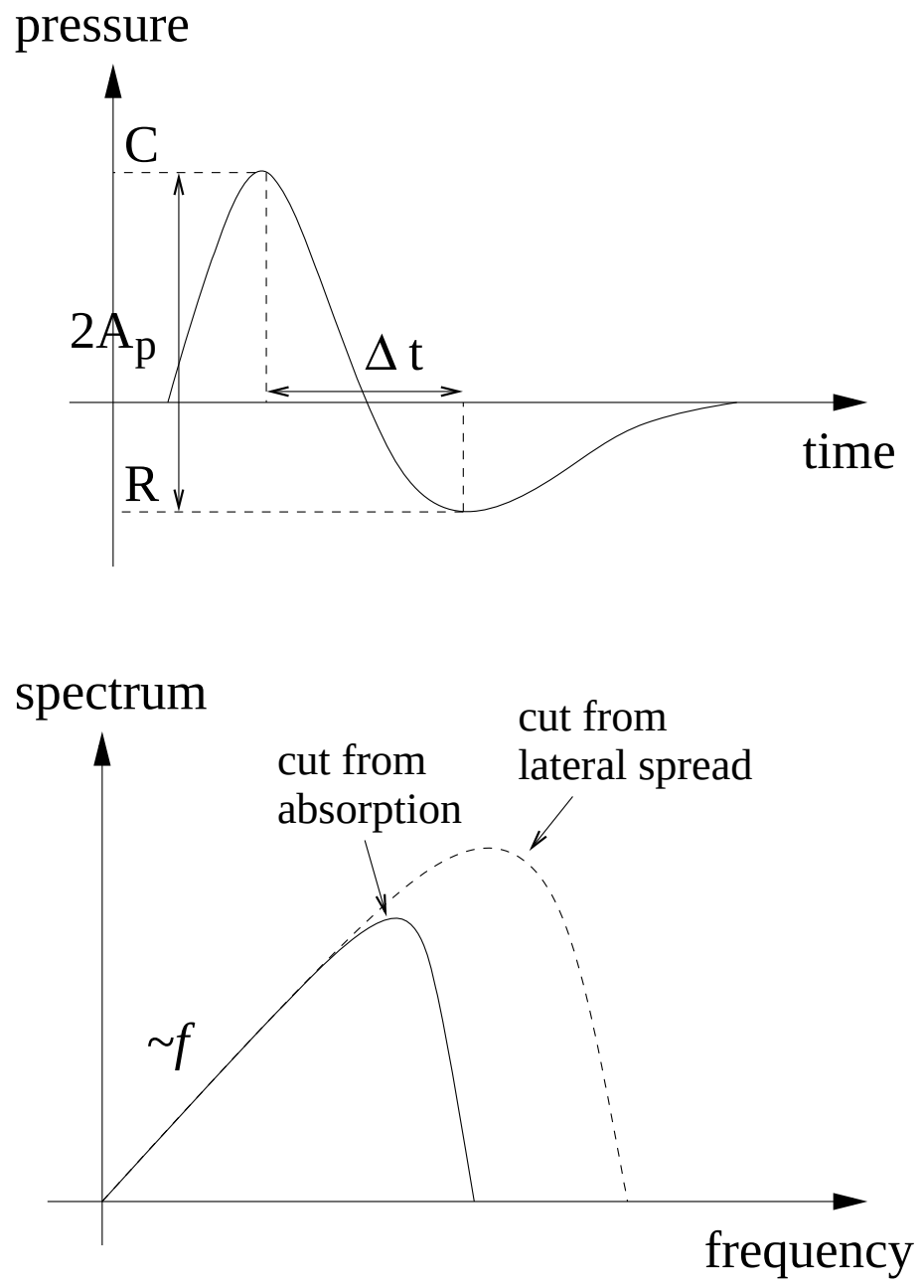

Figure 6: Sketch of a typical acoustic signal pulse. The upper plot represents the time pattern. The various characteristics discussed in the text are represented on this figure. The lower plot is the frequency spectrum. Depending on the distance to the source the high frequency content of the spectrum, resulting from the 'instantaneous' energy deposition, is suppressed at first by scattering from absorption or at first by the lateral spread of the energy deposition area.

**2-5-1) Near-field/far-field transition and directivity of the acoustic signal**

For a pure sine pressure wave of wavelength $\lambda$, far-field or Fraunhofer diffraction conditions are reached when the observation range is large compared to $r^* = L^2/(2\lambda)$, with $L$ the characteristic extension of the source. For cascades, because the source is coherent, but extends from tens to hundreds of metres, acoustic near-field conditions can be maintained up to a hundred kilometres. Hence, in particular for extended LPM cascades, Fraunhofer diffraction approximation is not valid over practical observation ranges.

For compact cascades of hadronic origin, the strengthening of sound absorption at distances larger than a hundred metres results in a transition from near to far-field conditions, as the frequency content of the signal shifts to longer wavelengths. At close range, one has perfect near-field conditions. The acoustic signal propagates 'rigorously' orthogonally to the cascade axis, with a close to cylindrical wave-front. The amplitude of the pressure field, along the cascade direction, is an image of the longitudinal energy deposition, in a point to point relationship. As absorption is enhanced, above a hundred metres, this image becomes 'fuzzy' and de-focused. As can be seen in figure 7, at a distance of 2 km the wave-front becomes spherical, close to the depth of the maximal of energy deposition. It spreads out reaching a constant –3 dB half-angle aperture of $\Delta\theta_{3db} = 0.55^{O}$ at ranges larger than 10 km with rigorous far-field conditions.

For LPM extended cascades the situation is more complicated. Because both the length $L$ of the energy deposition and propagation range increase with energy, near-field conditions are maintained over all practical observation ranges. The wave-front exhibits a close to cylindrical shape and the amplitude is a fuzzy image of the longitudinal energy deposition. However, because the energy deposition exhibits local maxima, the wave-front can have local curvatures. Hence, at ranges of several kilometres the arrival time of the signal may fluctuate by some tens of $\mu$s over the extension of the pressure field.

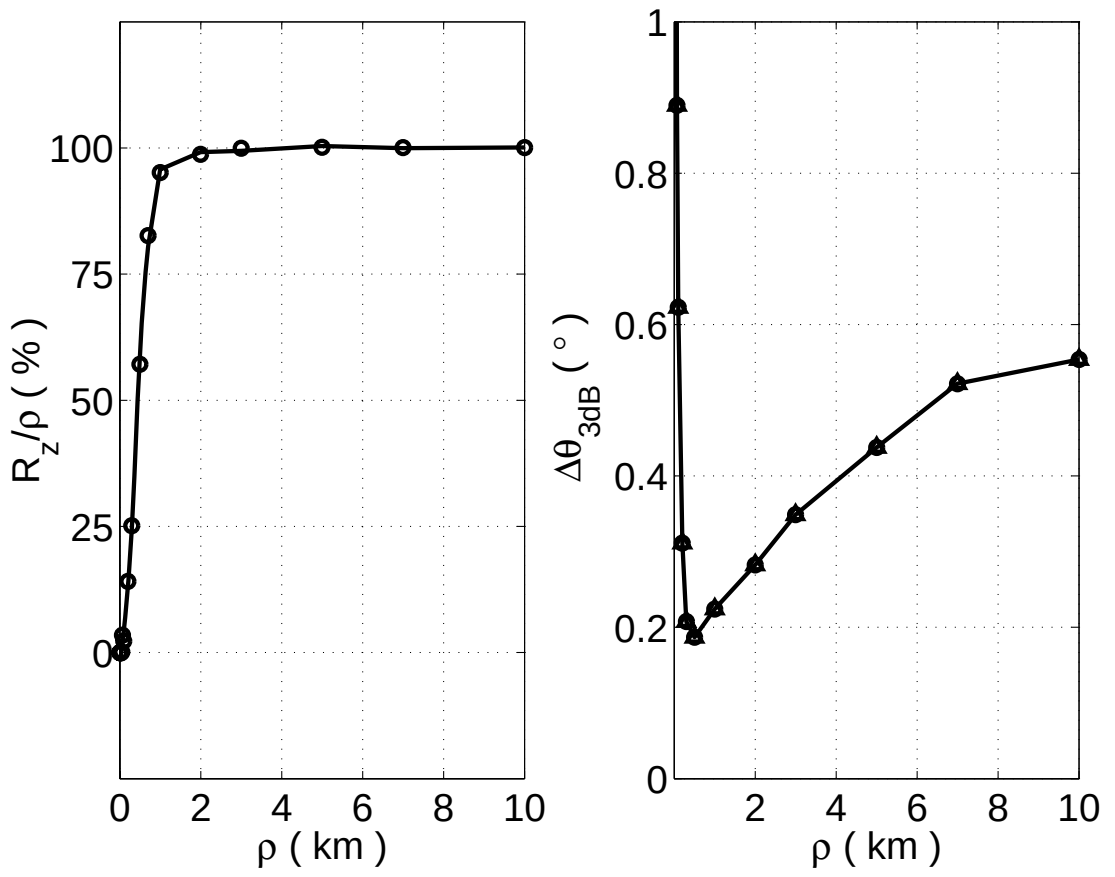

Figure 7: Far-field transition of the acoustic signal emitted by compact hadronic cascades ($10^{20}$ eV, NC). The plot on the left shows the curvature radius, $R_z$, normalised by range, $\rho$, as a function of this propagation range. The plot on the right is the half-angle angular aperture at –3 dB, $\Delta\theta_{3dB}$, also shown as a function of propagation range.

## 2-5-2) Amplitude of the acoustic signal and maximal propagation range

As was first pointed out by Learned [24], because of its broadband frequency content the amplitude of the acoustic signal only falls as a power-law with distance, not with an exponential cut-off. At very short ranges of some ten metres, the signal is in perfect near-field condition and absorption is negligible. As a result the amplitude of the signal falls as a square-root power-law with the distance, $r$, with an asymptotic behaviour as $1/\sqrt{r}$. At distances of some hundred metres, absorption becomes sensitive, bringing in another ½ power-law from Gaussian diffusion and the signal then falls as $1/r$. For compact hadronic cascades, at kilometric ranges, because of the spreading of the wave-front, the amplitude begins to fall faster reaching a $1/r^2$ power-law at 10 km. For the extended part of an LPM cascade, near-field conditions are maintained and the fall of amplitude with distance stays close to $1/r$.

From the computation of the acoustic signal at large distances one can deduce the maximal range, $\rho_{\max}$, for signal detection given an ambient noise level, $B$. At UHE energies, the maximal range, $\rho_{\max}$, can be parameterised as :

$$\rho_{\max} = \begin{cases} (545 \pm 45\ \text{m}) \left( \dfrac{E}{10^{18}\ \text{eV}} \right)^{0.56} \left( \dfrac{1\ \text{mPa}}{B} \right)^{0.70} & \text{(CC)} \\ (407 \pm 46\ \text{m}) \left( \dfrac{E}{10^{18}\ \text{eV}} \right)^{0.60} \left( \dfrac{1\ \text{mPa}}{B} \right)^{0.64} & \text{(NC)} \end{cases} \tag{13}$$

where $E$ is the energy of the incoming neutrino. This was found to reproduce the simulated data with about 10 % accuracy, for energies between $10^{18}$ and $10^{20}$ eV and with noise levels varying from 1 to 10 mPa. As a comparison, ambient acoustic noise for calm sea conditions (sea state 0) is about 3 mPa [26,27] in a practical frequency band for the signal of 100 kHz. The electrical thermal noise of a piezo ceramic detector, used for the transduction of a pressure wave in underwater acoustics, is several mPa [4] in the same 100 kHz frequency band.

## 2-5-3) Shape of the acoustic signal

Due to absorption, the shape of the signal varies significantly with the distance to the cascade. At short ranges of some tens of metres, the duration of the signal is dominated by the narrow width of the lateral energy distribution. The duration is very short, less than $\Delta t = 1\ \mu s$ with a MHz frequency content. Furthermore the signal is close to mono-polar with a symmetry ratio $R/C \approx 15\ \%$. At a hundred metres, due to the absorption transition, the duration of the signal suddenly increases by more than one order of magnitude, reaching $\Delta t = 10\ \mu s$. Simultaneously, the signal becomes even more mono-polar, with a symmetry ratio reaching a minimum value of about 5 %. This results from the 'tail' that the absorption response exhibits at these ranges. At larger ranges the duration of the signal is dominated by absorption. It increases as a square-root power-law with the range $r$, and its duration is close to those of the absorption impulse response, given by a diffusive law, as :

$$\Delta t = (18 \pm 3\ \mu s) \sqrt{\frac{1\ \text{km}}{r}} \tag{14}$$

In this regime, the duration of the signal gives an estimation of the distance to the impulse-like sound source. In addition, the signal becomes more and more bipolar, reaching a common symmetric shape at distances of tens of kilometres.

The combination of signal duration, $\Delta t$, and the $R/C$ ratio can be used to give some insight into the nature of a detected acoustic bipolar pulse. For a given compact cascade model, as range varies one describes a parametric curve in the plane defined by parameters $\Delta t$ and $R/C$, as can be seen on figure 8. At intermediary ranges, from a few hundred metres to a few kilometres, extended LPM cascades are characterised by a more asymmetric shape than compact hadronic ones. Therefore, LPM cascades are confined to the lower right part of this diagram. A detected bipolar pulse falling in the upper left part of the diagram would not be consistent with any cascade model, hence it is unlikely to be of cascade origin.

Nevertheless, such an estimator is dependant on the shape of the lateral distribution. Furthermore, it requires a broadband transducer device, with a steady phase response, not to alter the shape of the measured signal.

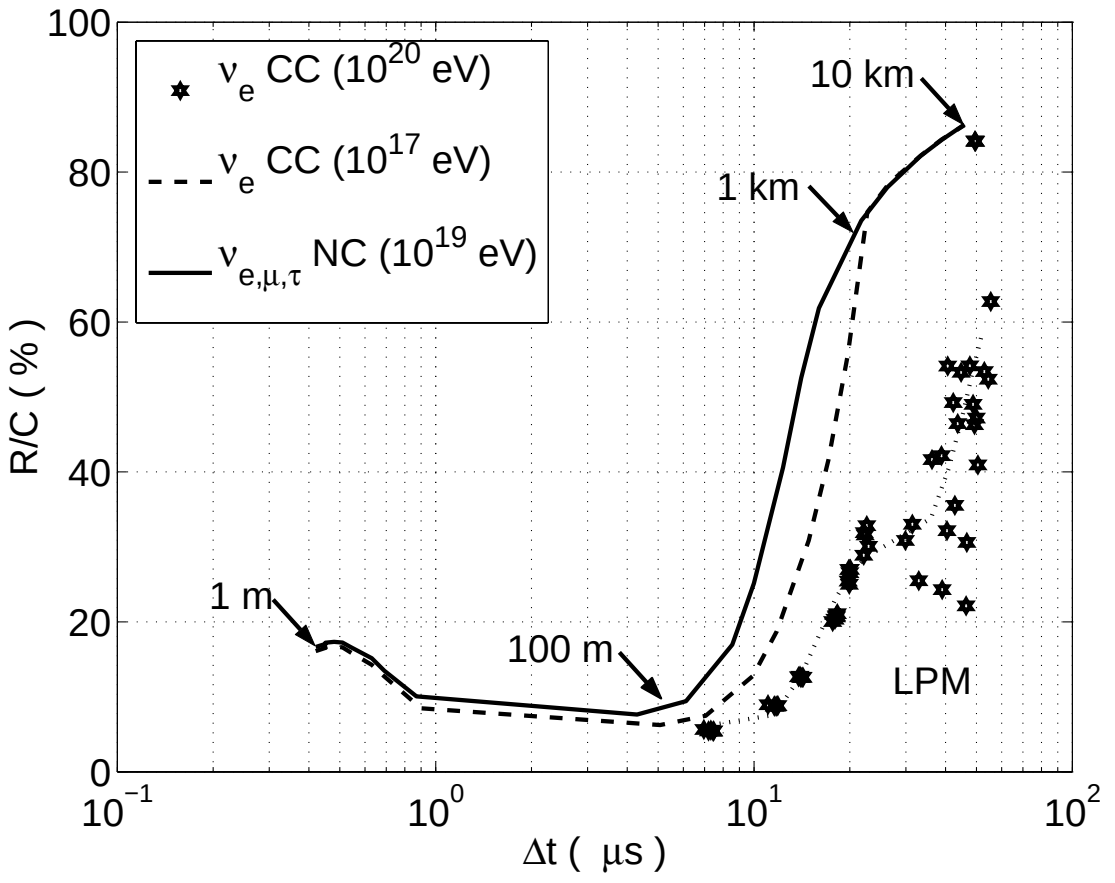

Figure 8: Characterisation of cascades thermo-acoustic signal by a *R/C* versus $\Delta t$ diagram. The range is indicated by arrows for the NC cascade model. The solid curve correspond to a $10^{19}$ eV NC neutrino interaction with solely a compact cascade of hadronic origin. The dashed curve is a $10^{17}$ eV CC $\nu_e$ interaction with a slightly elongated EM cascade due to LPM effect. Stars represent characteristics of some $10^{20}$ eV CC $\nu_e$ interactions with LPM strongly extended EM cascades.

### 2-5-5) Comparison to previous studies

The signal characteristics obtained in this work are compared to previous results in tables 1 and 2. Amplitude has been scaled according to the Gruneisen factor. We considered two cases: a compact hadronic cascade of 10 PeV and an extended LPM cascade of 100 EeV. Observation along the cascade direction is performed close to the depth of the maximum amplitude of the acoustic signal.

At a distance of 400 m from the cascade axis both absorption and the lateral distribution of energy deposition, $g_z$, are of importance for the signals shape. It can be seen from table 1 that previous works are in disagreement for both the amplitude and the shape of the signal. For hadronic cascades our amplitude and duration estimations are consistent with results of Dedenko *et al.* [10], however, we find a shape of the signal much more mono-polar, in better agreement with work of Learned [24].

At kilometric distances the shape of the signal is dominated by absorption. For LPM extended cascades, at a distance of 1 km from the cascade axis, we find reasonable agreement with Lehtinen *et al.* [12], thought signal duration could not be estimated from their work.

Table 1: Acoustic signal characteristics for a $10^{16}$ eV hadronic cascade at a distance of 400 m from the shower. Variations in amplitude are estimated from different energy deposition distributions, more or less compact.

| | $A_p$ (μPa) | $\Delta t$ (μs) | $R/C$ (%) |
|---|---|---|---|
| Askaryian [8] | 25 | 17 | 100 |
| Dedenko [10] | 44 | 10 | 75 |
| Learned [24] | 3 | 20 | 50 |
| This work | 47±5 | 10 | 35 |

Table 2: Acoustic signal characteristics for a $10^{20}$ eV LPM extended cascade at a distance of 1 km. Variations in amplitude results from the stochastic nature of the LPM energy deposition.

| | $A_p$ (mPa) | $\Delta t$ (μs) | $R/C$ (%) |
|---|---|---|---|
| Lehtinen [12] | 10 | ? | 23 |
| This work | 7±1 | 20 | 28-34 |

In addition, the fall in absorption for impulses at distances lower than hundred metres combined with a MHz frequency content of the source shower leads to opportunities for close range observation of 'low energy', PeV hadronic cascades, near the Glashow resonance [28]. At a distance of 40 m from the shower axis, for a hadronic cascade of 10 PeV we compute an amplitude of 1 mPa, with a duration of $\Delta t = 0.7$ μs and a ratio $R/C = 12\,\%$. At these distances the signal is in perfect near-field conditions and its shape is dominated by the lateral distribution of energy deposition, $g_z$. Therefore the observation of the acoustic pressure field would allow for imaging of the cascade energy distribution. This could find applications in particle physics. Nevertheless, specific transducers, sensitive to hundreds of kHz frequencies, are required to exploit this opportunity, and little is known about ambient acoustic noise in these frequency ranges.

## 3) EFFECTIVE VOLUME OF DETECTION AND FLUX SENSITIVITY LIMITS

### 3-1) Effective volume of detection for a single hydrophone in infinite extension water medium

The effective volume of detection, $V_{eff}$, for the acoustic signal from a cascade can be generally defined by an integral form as:

$$V_{eff}(\vec{u})=\int_V p_d(\vec{u},\vec{r})d^3\vec{r} \qquad (15)$$

where $p_d(\vec{u},\vec{r})$ is the probability to detect a cascade of orientation $\vec{u}$ and located in $\vec{r}$. The integration is performed over the whole water volume.

According to pressure field computation presented previously, we estimated the effective volume for a single hydrophone located in a water medium of infinite extension, with no boundaries. In this particular case the effective volume does not depend on the cascade orientation, $\vec{u}$. It was computed with Matlab [29] and interpolated with its default 'contour' algorithm. For the detection probability, $p_d$, we assumed a 'steep' law, with detection being certain ($p_d$=1) if the pressure level overcomes a given noise threshold *B*, and $p_d$=0 otherwise.

In the UHE energy range ($10^{18\text{-}20}$ eV) the effective volume can be modelled as follows For charged current $\nu_e$ interactions, because near field conditions are maintained over all observation ranges, the effective volume is in simple relation with the maximal propagation range, given by equation (13). The average effective volume goes as:

$$V_{eff}^{\infty} = \pi\rho_{\max}^2 L_{eff}$$
$$L_{eff} = (13.8 \pm 1.5 \text{ m})\left(\frac{1 \text{ mPa}}{B}\right)^{0.48}\left(\frac{E}{10^{18} \text{ eV}}\right)^{0.44} \quad \text{(CC)} \qquad (16)$$

which is the volume of a disk of radius $\rho_{\max}$ and with an effective width $L_{eff}$. Due to the stochastic nature of the extended LPM part of the cascade the volume can fluctuate by 10-20 %. One notices that the effective length of the signal, $L_{eff}$, increases with the energy close to a square root power law. This is consistent with the increase of the cascade length in the UHE LPM regime, as was seen in previous part 2-2-1.

For neutral current interactions, due to the transition from near field to far field conditions, the picture is more complicated. The effective volume was seen to be in 10-20 % agreement with the following model :

$$V_{eff}^{\infty} = \Delta\theta_{eff}\,\frac{4}{3}\pi\rho_{\max}^3 \qquad (17)$$
$$\Delta\theta_{eff} = (1.2 \pm 0.1)\Delta\theta_{3dB}(\rho_{\max}) \quad \text{(NC)}$$

which is the volume of spherical section of small angular half aperture, $\Delta\theta_{eff}$, as sketched on figure 9. The effective aperture, $\Delta\theta_{eff}$, varies with the maximal range, $\rho_{\max}$. It is nearly proportional to the -3 dB angular aperture, previously shown on figure 7.

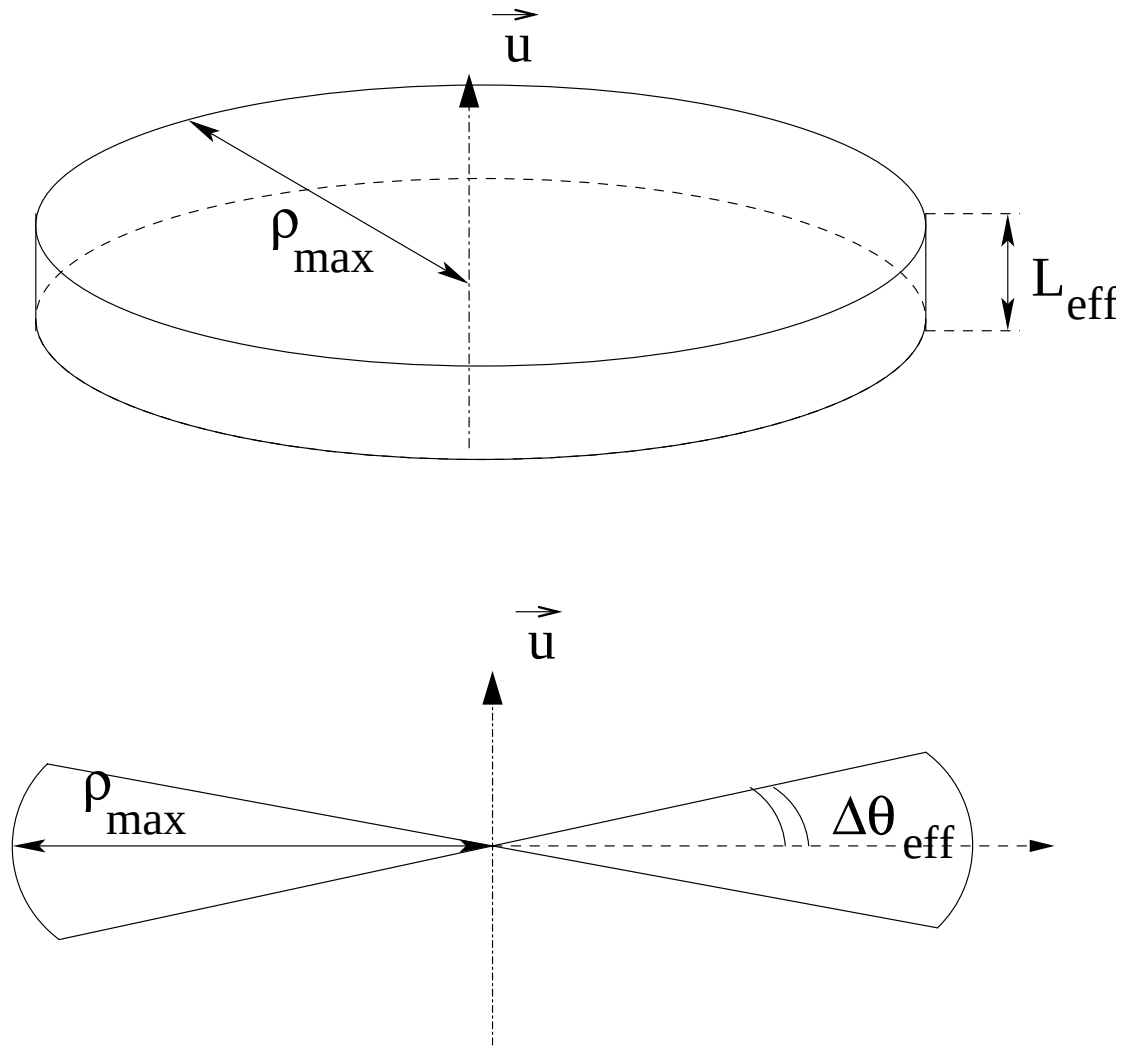

Figure 9: Modelling of the wave-front. The upper plot is a disk like model, with perfect near-field conditions. The lower plot is a far-field model, with a spherical curvature of the wave-front, and a small angular aperture.

### 3-2) Surface and bottom boundaries

When considering propagation ranges of several kilometres or more it is important to keep in mind that our water medium has boundaries. In particular the depth of the sea is limited, to 2.5 km on the ANTARES site [5]. Due to the strong density mismatch, the water to air boundary is a perfect reflector, with a 180° phase shift, however, the sea surface has a rugosity and is screened by a layer of bubbles. This results in scattering of the acoustic signal, and a loss of coherency, hence in a weaker amplitude. Furthermore, at a grazing angle the 180° phase shift results in destructive interferences. The bottom boundary is even more complex, with transmission, reflection and reverberation factors, which vary depending on the composition of the sea bed. Therefore, a direct detection, with no boundary issues, appears as the most promising for the detection of our acoustic signal. In particular at very long ranges where the signal may undergo multiple reflections, efficient propagation is limited to nearly horizontal directions.

In order to quantify these limitations we computed the efficiency for direct detection, $\varepsilon_{eff}$, defined as the ratio of the mean effective volume with boundaries, to the effective volume without boundaries, as :

$$\varepsilon_{eff} = \frac{<V_{eff}>}{V_{eff}^{\infty}} \leq 1 \qquad (18)$$

where the mean value is computed over all orientations, $\vec{u}$, of cascades. For the detection probability, $p_d$, we considered two limit cases, sketched on figure 9. The first one is an ideal near field wave front, with $p_d = 1$ within a cylindrical volume of radius $\rho_{\max}$ and height $L_{eff}$, and $p_d = 0$ outside. The second model is ideal far-field with a spherical volume of half angular aperture $\Delta\theta$. For the cascade orientations, $\vec{u}$, we assumed a uniform distribution over a half sky coverage. The full details of the computation are given in Appendix L of [4]. In the computation, we further assume that the signal is directive so that one can neglect corrections to the efficiency of order $L_{eff} / \rho_{\max}$ or $\Delta\theta_{eff}$.

We found that under the assumption of direct detection and for a focused signal, the efficiency is given as :

$$\varepsilon_{eff}^{1H} = 1 - \frac{1}{2} J(\frac{z}{\rho_{\max}}) - \frac{1}{2} J(\frac{H-z}{\rho_{\max}}) \qquad (19)$$

where $z$ is the depth of the hydrophone, and, $H$, the total height of water. The loss function, $J$, depends on the wave-front model. It goes as :

$$J(x) = \begin{cases} (1-x)^2 & \text{(near-field)} \\ (1-x)^2(1+\frac{1}{2}x) & \text{(far-field)} \end{cases} \quad \text{if } x \leq 1$$
$$J(x) = 0 \quad \text{otherwise} \qquad (20)$$

The efficiency is maximal for a hydrophone located at mid-depth, neglecting refraction effects, and falls at long ranges as $\varepsilon_{eff}^{1H} \propto H / \rho_{\max}$. Hence at long ranges as a function of the water height it is important to take into account the limited extension of the available water volume when estimating the effective detection volume.

In addition, due to the strong directivity of the signal, the efficiency loss does not affect vertical cascades, but mostly horizontal ones. This is seen as a fall in geometric efficiency as the angle, $\theta$, that the cascade makes with the vertical increases. The loss of efficiency with cascade orientation can be characterised by a 3 dB angular aperture, $\Delta\theta_H$. At ranges, $\rho_{\max}$, larger than the water height, $H$, the angular aperture little depends of the wave-front model. It goes as :

$$\Delta\theta_H \approx \frac{H}{2\rho_{\max}} \text{ (rad)}, \quad \rho_{\max} \geq H \qquad (21)$$

Therefore, for kilometric deep water layers, at ranges of tens of kilometres, detection is limited to vertical cascades within a few degrees angular aperture. Hence, the medium naturally selects vertical cascades.

### 3-3) Refraction and shadowing from the sea bed

In the Mediterranean Sea the temperature is stable, close to 13ºC below a hundred metres depth. As a result, sound velocity varies linearly with depth, with a gradient of $k_c = 1.65 \text{ cm} \cdot \text{s}^{-1} \cdot \text{m}^{-1}$, due solely to pressure variations. Consequently, acoustic rays are bent along circular trajectories in the vertical dimension with a characteristic curvature radius of $R_c \approx 90 \text{ km}$. The directivity of the acoustic signal from a cascade is affected by this curved ray structure. The most important effect of refraction is that the wave-front of the signal is deflected toward the surface. At a distance $\rho$ from the shower axis, the wave-front is rotated from the orthogonal direction by an angle $\theta_r$ resulting in a deflection $\Delta D$ along the shower axis direction. For small angles, the angular shift and the deflection can be approximated as :

$$\theta_r = \frac{\rho}{R_1} \quad \text{and} \quad \Delta D = \frac{\rho^2}{2R_1} \qquad (22)$$

where $R_1$ is an effective curvature radius, depending on the angle $\theta$ that the shower makes with the vertical direction, as $R_1 = R_c / \cos(\theta)$. The arrival time pattern is affected in a more complex manner, as can be inferred from equation (31) in appendix B. In particular, arrival times are no longer symmetric by rotation around the shower axis. Nevertheless, locally the characteristics of the signal, apart from an angular shift, are little affected by refraction.

Due to the strong directivity of our signal, refraction has little effect on the efficiency of detection, in an infinite medium. This can be understood as follow. For each cascade signal that misses our hydrophone because of refraction, one can find another cascade that is detected due to refraction, but would not have been detected otherwise.

However, for a linear sound velocity profile, as in the Mediterranean Sea, with acoustic rays being deflected toward the surface, the bottom boundary introduces a bias. For a given source, the sea bed casts an acoustic shadow delimited by the ground grazing acoustic ray. A general scheme is given in [4,26]. The water volume located farther away than this grazing ray is out of reach, as it would demand rays to pass through the bottom. Consequently, at long ranges, when the deflection, $\Delta D$, of rays becomes similar to the full water height, *H*, direct detection is no longer possible. For example, for $H = 2.5$ km this sets a limit at some 20 km range. Hence, due to refraction the total volume of water available for direct detection is limited. This results in a fall in efficiency at large ranges, reflected by a shadow factor *F*, defined as :

$$\varepsilon_{eff}^{R} = (1 - F)\varepsilon_{eff} \tag{23}$$

where $\varepsilon_{eff}^{R}$ is the total efficiency taking into account surface-bottom limitations and bottom shadowing from refraction, and $\varepsilon_{eff}$ is the efficiency without refraction. Analytical computation of this shadow factor is not straight forward, therefore the integral (15) for the effective volume, taking into account shadowing from the sea bed and refraction, was estimated by Monte-Carlo. The details of the computation can be found in Appendix S of [4]. A numerical estimation of the shadow factor is given on figure 10 for a 2.5 km water depth and for an hydrophone located 450 m above the sea bed. This situation is similar to an ANTARES line [5], with an hydrophone fixed close to the top of the line.

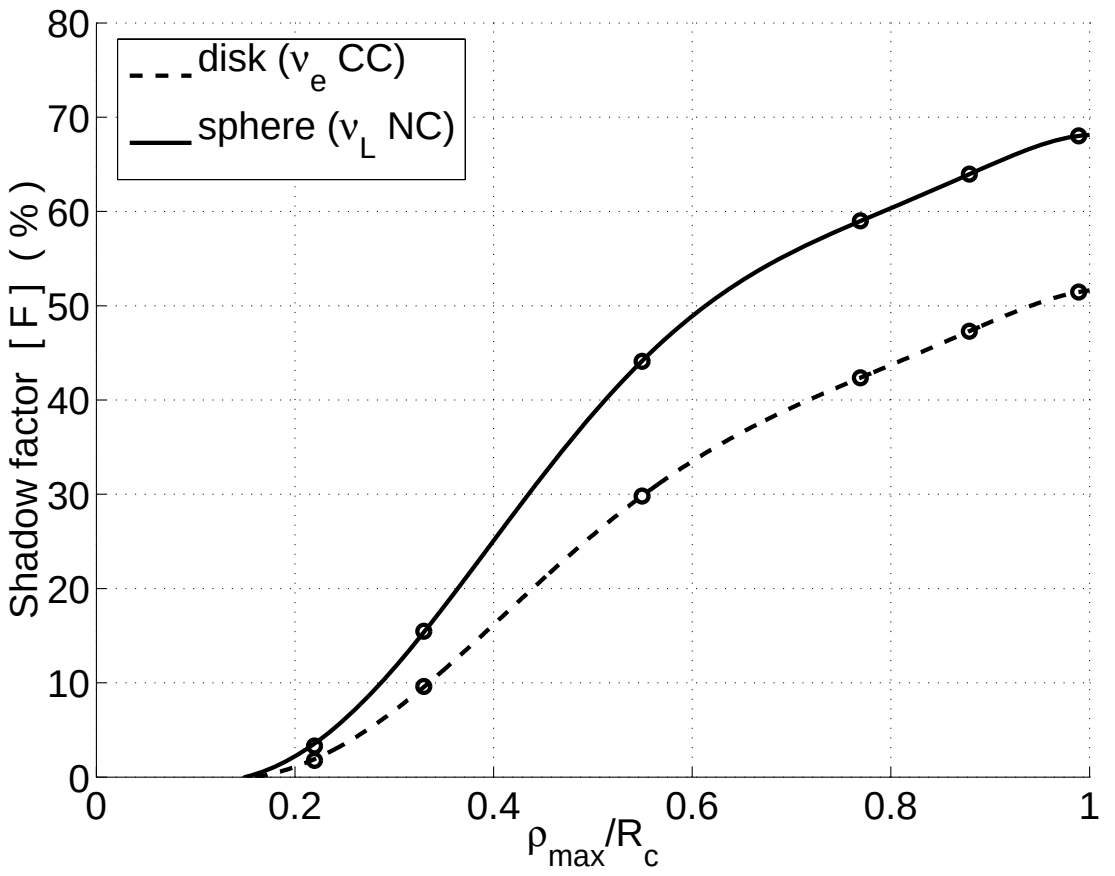

Figure 10: Shadow factor *F* by the sea bed as a function of maximal range. Range is normalised by refraction curvature radius $R_c \approx 90$ km. The hydrophone is located 450 m above the sea bed, for a full water height, *H*, of 2.5 km.

One may think of using refraction, to our advantage, with deep sea oceanic channels, in order to survey larger water volumes. However, on larger scales, sound propagation will still be limited by horizontal non-homogeneities. An exponential cut-off, from chaotic perturbations, is to be expected at distances of a hundred kilometres [30]. Furthermore, oceanic channels are 'cold' water at 1-2 km depth only, where the thermo-acoustic conversion is not the most efficient.

### 3-4) The independent hydrophones limit for multiple sensors.

For a set of *N* hydrophones, the probability, $p_d$, to detect a given cascade signal, with one or more hydrophones, is bounded by :

$$p_d \leq \sum_{i=1}^{N} p_{d,i} \tag{24}$$

where $p_{d,i}$ denotes the probability to detect the cascade with the i$^{th}$ hydrophone. The equality in previous equation (24) is achieved only for 'independent' hydrophones, which means here : hydrophones 'surveying' different water volumes, with a null intersection. Assuming hydrophones are all identical and located at similar depths, it follows from the inequality (24) that the effective volume of detection for a set of *N* hydrophones, $V_{eff}^{NH}$, is bounded by the volume for a single hydrophone, $V_{eff}$, as :

$$V_{eff}^{NH} \leq N V_{eff} \tag{25}$$

The reconstruction of cascade events requires multiple hydrophones to detect a same signal. In addition the strong directivity of the signal implies a dense arrangement of hydrophones in order to sample the acoustic wave-front. The detection of events in coincidence on multiple hydrophones is not limited by the range of the signal, of ~1 km, but by the extension of the wave-front, which is of $L_{eff} \approx 10 - 100$ m. Consequently, in practical situations the effective volume of detection for a set of *N* hydrophones is lower by one or two orders of magnitude than the independent hydrophones limit. A morecomplete discussion on the efficiency loss resulting form multiple hydrophones used in coincidences, such as compact structures, is beyond the scope of this report. but can be found in chapter 3 of [4].

### 3-5) Sensitivity to an Astrophysical flux

The sensitivity to astrophysical fluxes of a single hydrophone, was computed according to effective volume given by equations (16) and (17) in section 3-1, and extrapolated to extreme energies. We took into account limitations from the bottom, the surface and shadowing from refraction. We considered a hydrophone located at 450 m above the sea bed, with a full water height of 2.5 km. We used Standard Model UHE neutrino cross section, modelled according to Gandhi *et al.*[2], and a $2\pi$ sky coverage. The sensitivity was computed for noise levels varying from 1 to 10 mPa, which are representative of ambient acoustic sea noise and instrumentation noise.

Table 3: Sensitivity limit of a single hydrophone to a $1/E^2$ flux for various signal amplitudes. Limits are given with a 90% confidence bell, according to Feldman and Cousins [35], and for 1 year of observation. Units for flux limits are $\mathrm{GeV \cdot cm^{-2} \cdot sr^{-1} \cdot s^{-1}}$

| Noise threshold → | 1 mPa | 3 mPa | 10 mPa |
|---|---|---|---|
| $\nu_e$ (CC) | $4.3 \cdot 10^{-7}$ | $1.5 \cdot 10^{-6}$ | $6.2 \cdot 10^{-6}$ |
| $\nu_{\mu,\tau}$ (CC), $\nu_L$ (NC) | $1.0 \cdot 10^{-6}$ | $3.3 \cdot 10^{-6}$ | $1.4 \cdot 10^{-5}$ |

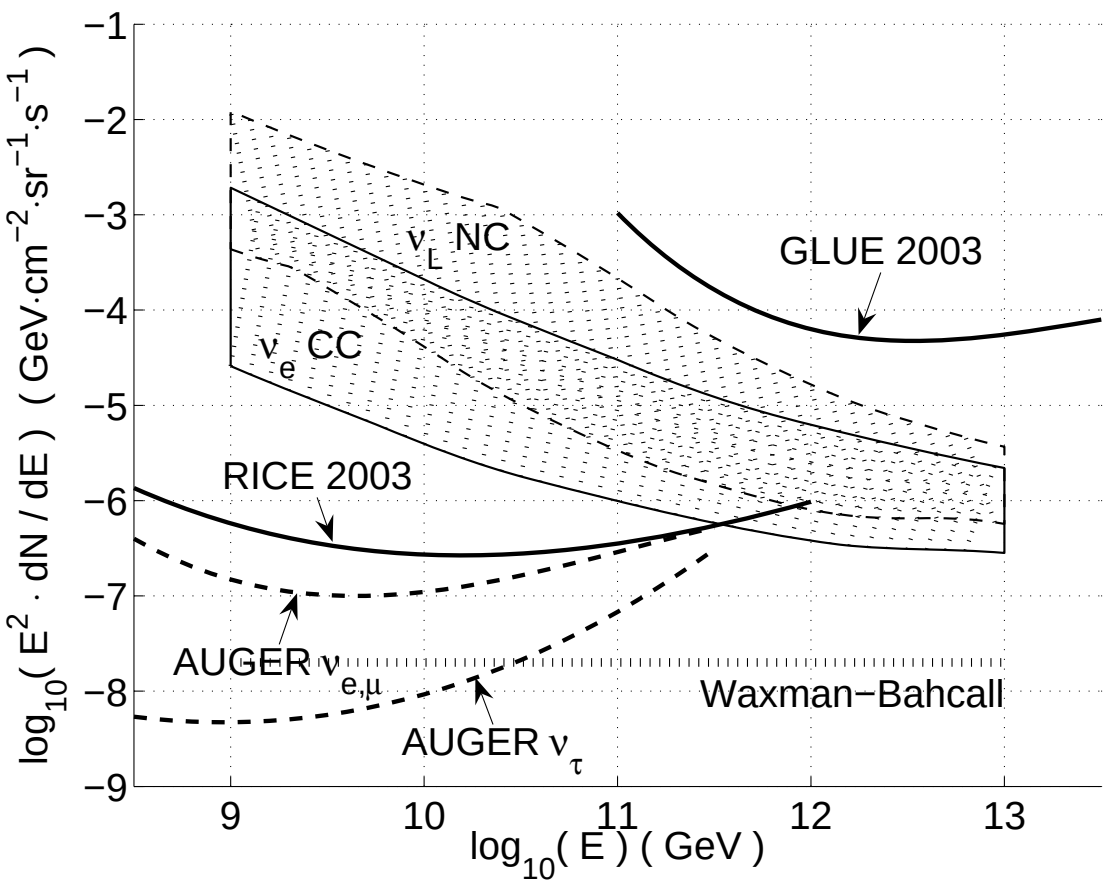

Figure 11: Sensitivity limits for a single hydrophone. Limit is given for 1 year of observation and with the assumption of 1 event detected per decade in energy. The dashed boxes indicate the results of this study, for a $\nu_e$ charged current interaction (solid contour) or a neutral current interaction (dashed contour), and for various noise threshold levels from 1 to 10 mPa. Ambient acoustic noise originates from the sea surface. It is expected to be of 3 mPa for a calm sea state, and so to reduce with depth. Though, this latter point is not firmly established [26].

Results for two cases are shown on figure 11 : for a 'golden' $\nu_e$ charged current event, and for a solely hadronic core. Sensitivity limits for this work, and for the AUGER [31] experiment are shown with the assumption of a density of detected events of 1 per decade in energy and for 1 year of observation. We also show the RICE [32] and GLUE [33] experimental limits, for purpose of comparison. The Waxman-Bahcall [34] upper bound is indicative of the expected level of potential neutrino fluxes. In table 3, we also give the sensitivity limits to a $1/E^2$ flux, extending on energies from $10^{18}$ to $10^{22}$ eV. The sensitivity limit were computed from the Feldman and Cousins [35] confidence bell method with the assumption of no neutrino candidate event. The limits obtained are consistent with results shown on figure 11. They are dominated by the highest energy particles considered, close to the flux upper limit at $10^{22}$ eV.

At $10^{18}$ eV, for the most optimistic case of $\nu_e$ CC interactions with low noise levels of 1 mPa, the flux sensitivity limit of a single hydrophone is of $E^2\phi_0 = 3 \cdot 10^{-4}\ \mathrm{GeV \cdot cm^{-2} \cdot sr^{-1} \cdot s^{-1}}$. Therefore, to achieve the experimental limit of RICE 2003, one needs a hundred independent hydrophones. Competing with the AUGER $\nu_{e,\mu}$ limit requires several hundred independent hydrophones. In what we think a more realistic situation, with a 3 mPa noise level and redundancy among sensors, one would need thousands to tens of thousands hydrophones to compete already existing or near future experimental limits. However, one sees that the sensitivity improves with increasing energy, even when taking into account boundary limitations and refraction.

For the location studied here, the sensitivity limit would flatten around $10^{22}$ eV giving, for a single hydrophone, a flux sensitivity limit of $E^2\phi_0 \approx 3 \cdot 10^{-7}\ \mathrm{GeV \cdot cm^{-2} \cdot sr^{-1} \cdot s^{-1}}$. Therefore, it seems to us, that with a limited instrumentation it is only at extreme energies, above $10^{20}$ eV, that the acoustic method can become really efficient. By using tens of independent antenna arrays, surveying complementary water volumes, the sensitivity to extreme energy neutrino fluxes can be lowered to the Waxman-Bahcall limit. At these extreme energies the propagation ranges to tens of kilometres and one really benefits from the large available water volume. Consequently, boundaries are very important, and it is crucial for this purpose of extreme energies, to locate the experiment in a deep sea, off-shore area. Nevertheless, a point of caution should be added here since as the propagation range increases the signal shifts to a lower frequency content, of some kHz, where ambient acoustic noise is higher.

We finally point out that by extrapolating this scheme to very extreme energies, at some $10^{23\text{-}24}$ eV, in the narrow core density of the cascade, the energy deposition density would allow water evaporation. Hence, competitive ablation mechanisms are to be expected, that may start from $10^{22}$ eV. Such mechanisms are potentially more efficient than the thermo-acoustic one.

## 4) CONCLUSIONS

Previous calculations on the acoustic signals from hadronic cascades show significant disagreement on both the amplitude and the shape of the signal. The disagreement can be related to the different sound absorption and energy deposition models used. In this paper, the acoustic signal has been evaluated using a more realistic power-law lateral distribution of the enrgy deposition, and a more complete absorption model, including phase effects. At a distance of 400 m from the cascade axis we find an enhanced pressure level, in agreement with work of Dedenko *et al.* [10] but following Learned [24] we predict an asymmetric shape for the acoustic signal. At kilometric ranges the signal is dominated by absorption and our results for LPM extended cascades are in agreement with Lehtinen *et al.* [12]. At distances below a hundred metres, due to the fall in absorption and the narrow lateral distribution, we compute an enhanced amplitude with a MHz frequency content and a close to mono-polar signal shape.

At these short ranges, the acoustic pressure field is in perfect near field conditions, and the shape of the acoustic impulse strongly depends on the shape of the lateral energy distribution. Consequently, at close range, hydrophones can accurately image the cascade energy deposition. This might find applications on giant cubic kilometres Cerenkov neutrino detectors. A dense lattice of acoustic sensors could complement the optical detector for imaging of PeV cascades.

In this work it is shown that the acoustic technique becomes really efficient at extreme energies, above $10^{20}$ eV. At UHE, the kilometric range of the signal is not sufficient to detect fluxes close to the Waxman-Bahcall limit, with a reduced instrumentation. At extreme energies propagation ranges to several tens of kilometres and boundaries are important. Furthermore, though the Mediterranean Sea is a favourable medium for the thermo-acoustic generation mechanism, the particular refraction patterns limits direct detection to about 20 km. Nevertheless, if sources of neutrinos with such extreme energy exists, from tens of independent antenna arrays, fluxes could be observed close to the Waxman-Bahcall limit. In addition, at these extreme energies an enhanced signal amplitude can be expected from both the cascade development, where competitive processes could slow down the LPM extent of the shower, and from ablative sound generation mechanisms competing the thermo-acoustic effect.

### Acknowledgments

We thank the ANTARES collaboration for the opportunity and the support with this work. For discussions and advices with LPM shower simulations, we thank Leonid Dedenko. For support with English, and helpful criticisms, we thank John Carr and an anonymous referee.

### Appendix A : EM Cascade simulation in the deep LPM regime

UHE showers of electromagnetic origin are simulated using a two step scheme. A basic assumption in shower simulations is that 'actor' particles: $e^+$, $e^-$ and $\gamma$ do not interact each with another, but only with 'spectators', or matter. Consequently, at a given energy scale any shower can be considered as a stochastic sum of secondary showers.

The second point is that the stochastic behaviour of showers can be schematically understood as resulting from two factors: local statistic fluctuations, that are weakly correlated along the shower depth and follow the central limit theorem, and a global fluctuation of the shape of the shower due to the quantic fluctuations of the very first interactions. Local fluctuations decrease with the density of actor particles down to a few percent at PeV energies. Statistical fluctuations of primary interactions are an 'irreducible' cause of stochasticity which becomes predominant in the deep LPM regime.

Following these considerations, we first simulate the cascade high energy part by Monte-Carlo methods, down to an energy threshold, $E_{death}$, and then build the shower with deterministic parameterisations for secondary showers generated by particles with energies below $E_{death}$. For the Monte-Carlo step, with a threshold $E_{death}$ far above the critical energy ($\mathrm{E_c} = 54.3\ \mathrm{MeV}$ in water), only two processes of importance are considered: Bremsstrahlung and pair creation. Because of the high energy involved, these particles are perfectly focused along the direction (*Oz*) of the very first primary and a 1-dimensionnal Monte-Carlo is used.

Cross sections for Bremsstrahlung and pair creation processes are modelled according to Migdal [14]. The total cross section of these processes provides the interaction length, $L_{\mathrm{int}}$, allowing for randomisation of the travel path of an actor particle, 'before it interacts'. At the interaction point, the relative energy repartition, *x,* between secondary particles is randomised by a Metropolis algorithm. A Markovian chain $(x_n)$ is generated from two independent chains of random numbers, $(u_n)$ and $(v_n)$, uniformly distributed over [0;1]. Two chains of $10^3$ uniform random numbers are enough to provide *x* with a density reliable over more than 6 order of magnitudes.

For the second step, the reconstruction of the shower, particles with energies below $E_{death}$ are binned : according to their nature, to the logarithm of their energy and to their 'birth' location along the (*Oz*) axis. The shower is reconstructed from this discrete binning with a Finite Impulse Response (FIR) algorithm, where the FIR is a 2-dimensionnal ($\rho, z$) parameterisation of the shower. The shower parameterisations are first issued by GEANT4, for primary energies between $10^{14}$ eV and $10^{17}$ eV, then parameterised and the resulting showers extend the procedure to the UHE range.

The whole code was developed with Matlab [29] and compiled in C to speed up the execution time. The compiled code was run on the IN2P3 PC farm [36], at Lyon. The first Monte-Carlo step takes a few minutes on a single Pentium III of the farm (1.4 GHz, 2 GO RAM). The reconstruction, however, can vary from some minutes to 1 hour at UHE, as the energy gets diluted over space. We generated about two thousand LPM showers, with energies ranging from $10^{14}$ eV to $10^{20}$ eV.

**Appendix B : Acoustic signal computation**

The 3-dimensional integral given by equation (10) is reduced to a 1-dimensional integral by exploiting the axial symmetry and causality property. Reduction is achieved by the transformation of the integration over the cascade depth, *z*, to integration over the travel time given by a ray model. Details of computation can be found in Appendix H of [4]:

In a cylindrical set of coordinates ($\rho,\varphi,z$) where the (*Oz*) axis is defined by the cascade direction, the pressure field, *p*, is given by a sum of two 'sonic' rays originating from the cascade and arriving at coincident times. It goes as :

$$p(\rho,z,\varphi,t) = \frac{\alpha}{4\pi C_p}\frac{\partial}{\partial t}\sum_{i=1}^{2}\int_0^{\pi/2} f_z(z_i)G_z(\rho-\rho_0,z_i)d\phi \qquad (26)$$

with $G_z$ a distribution of the lateral energy distribution, $g_z$, defined as :

$$G_z(\rho,u) = \frac{1}{\pi}\int_{\rho'=|\rho|}^{+\infty}\frac{\rho' g_z(\rho',u)}{\sqrt{\rho'^2-x^2}}d\rho' \qquad (27)$$

The transform $G_z$ depends only on the lateral distribution of energy along the cascade. For a LPM cascade this latest is approximatively constant along the shower depth. For compact cascades it follows a scaling law as $u = z/z_{\max}$, where $z_{\max}$ is the depth of maximum density. Hence, $G_z$ only depends on the dimensionless parameter *u*. A semilog-polynomial law was used to fit $G_z$ over the range $\rho\in[0.2;800]$ mm and $u\in[0.5;2]$, with a 5% accuracy. The parametrisation was extended as constant beyond this range. $G_z$ behaves roughly as the shower lateral cross section, as $G_z(\rho,u)\approx 2\pi\rho g_z(\rho,u)$. More quantitative data are given in Appendix J of [4].

Parameters, $z_i$, and, $\rho_0$, in equation (26) depend on the observation location, time, *t*, and on the acoustic rays structure. For a constant sound velocity, $c_s$, they are given as :

$$\begin{cases}\rho_0 = c_s t\sin(v)\\ z_i = z \pm c_s t\cos(v)\end{cases} \qquad (28)$$

Hence, the pressure field results from two symmetric 'sonic' rays, originating from above ($z_i \geq z$) and below ($z_i \leq z$) the observation point. For a linear sound velocity profile, of gradient $k_c = |\vec{\nabla}c_s|$, expressions are more complex. Neglecting (de)-focusing effects parameters, $z_i$, and, $\rho_0$ are approximated as :

$$\begin{cases}\rho_0 = \rho_1(\beta\sin(v))\\ z_i = z \pm \frac{R_c - z_c}{2}\cos(\theta)\left[\beta\sqrt{\beta^2+\beta_1^2}\cos(v)-\beta^2\right]\end{cases} \qquad (29)$$

where $\theta$ is the angle that the shower makes with the vertical and :

$$\begin{cases}\beta^2 = 2(\cosh(k_c t)-1)\\ z_c = z\cos(\theta)-\rho\cos(\varphi)\sin(\theta)\end{cases} \qquad (30)$$

with :

$$\rho_1(x) = \frac{R_c - z\cos(\theta)}{1-\frac{x^4}{4}\cos^2(\theta)\sin^2(\theta)\cos^2(\varphi)}\{\sin(\theta)\cos(\varphi)[\ \ldots$$
$$\frac{x^4}{4}\cos^2(\theta)+\frac{x^2}{2}]+\sqrt{\frac{x^4}{4}[1-\sin^2(\theta)\sin^2(\varphi)]+x^2}\ \} \qquad (31)$$

From previous equations (29-31) ones sees that, except for a vertical cascade, in a vertical sound velocity gradient the pressure field is no more symmetric by rotation around the cascade axis. Nevertheless, practically the directivity pattern of the signal depends little on $\varphi$ but mainly on $\theta$.

In addition, since values of the angular parameter $\phi$, in equation (29), characterise the aperture of the acoustic signal, one can limit the numerical integration to small angles, close to $\pi/2$